\documentclass[12pt,a4paper]{article}
\usepackage{amsmath,amssymb,amsthm} 
\theoremstyle{remark}

\newcommand{\beq}{\begin{equation}}
\newcommand{\eeq}{\end{equation}}
\newcommand{\beqnn}{\begin{equation*}}
\newcommand{\eeqnn}{\end{equation*}}
\newcommand{\rd}{\partial}

\newcommand{\CC}{\mathbb{C}}
\newcommand{\PP}{\mathbb{P}}
\newcommand{\RR}{\mathbb{R}}
\newcommand{\ZZ}{\mathbb{Z}}
\newcommand{\calP}{\mathcal{P}}
\newcommand{\bsc}{\boldsymbol{c}}
\newcommand{\bst}{\boldsymbol{t}}
\newcommand{\bsx}{\boldsymbol{x}}
\newcommand{\bszero}{\boldsymbol{0}}
\newcommand{\frakL}{\mathfrak{L}}

\begin{document}

\title{Hurwitz numbers and integrable hierarchy of Volterra type}
\author{Kanehisa Takasaki\thanks{E-mail: takasaki@math.kindai.ac.jp}\\
{\normalsize Department of Mathematics, Kindai University}\\ 
{\normalsize 3-4-1 Kowakae, Higashi-Osaka, Osaka 577-8502, Japan}}
\date{}
\maketitle

\begin{abstract}
A generating function of the single Hurwitz numbers of 
the Riemann sphere $\mathbb{CP}^1$ is a tau function 
of the lattice KP hierarchy. The associated Lax operator $L$ 
turns out to be expressed as $L = e^{\mathfrak{L}}$, where 
$\mathfrak{L}$ is a difference-differential operator 
of the form $\mathfrak{L} = \partial_s - ve^{-\partial_s}$. 
$\mathfrak{L}$ satisfies a set of Lax equations that form 
a continuum version of the Bogoyavlensky-Itoh (aka hungry Lotka-Volterra) 
hierarchies.  Emergence of this underlying integrable structure 
is further explained in the language of generalized string equations 
for the Lax and Orlov-Schulman operators of the 2D Toda hierarchy. 
This leads to logarithmic string equations, which are confirmed 
with the help of a factorization problem of operators.

\end{abstract}

\begin{flushleft}

\end{flushleft}

\newpage

\section{Introduction}

The Hurwitz numbers of the Riemann sphere $\CC\PP^1$ count 
inequivalent finite ramified coverings of $\CC\PP^1$ 
by compact Riemann surfaces.  Okounkov considered 
a particular set of Hurwitz numbers, called 
{\it the double Hurwitz numbers\/}, and observed 
that a generating function of the double Hurwitz numbers 
is a tau function of the 2D Toda hierarchy \cite{Okounkov00}.  
This generating function can be specialized 
to a generating function of {\it the single Hurwitz numbers\/}. 
The specialized generating function becomes a tau function 
of the KP hierarchy \cite{GJ08,MM08,Kazarian08}. 
The integrable structures of the double Hurwitz numbers 
are further studied from various aspects 
\cite{Alexandrov11,Takasaki12,AMMN12,AMMN14,HO15}. 

We now reconsider the single Hurwitz numbers, and point out that 
a more fundamental integrable hierarchy is hidden therein.  
This integrable hierarchy is a large-$p$ (or {\it continuum}) limit 
\cite{Bog88,Itoh88} of the $p$-step Bogoyavlensky-Itoh 
(aka hungry Lotka-Volterra) hierarchy \cite{Bog87,Itoh87}. 
The lowest equation of this hierarchy 
is a 2D Toda-like field equation.  
Okounkov and Pandharipande remarked, 
in the course of studies on the Gromov-Witten theory 
of $\CC\PP^1$, that a generating function 
of the single Hurwitz numbers satisfies 
this Toda-like equation \cite{OP01}.  
We show how this equation emerges in the Lax formalism 
of the 2D Toda hierarchy. 

The ``continuum'' version of the Bogoyavlensky-Itoh hierarchies 
has a Lax operator of the somewhat unusual form 
$\frakL = \rd_s - ve^{-\rd_s}$, where $v$ is a function 
of $s$ and time variables.   
This is a linear combination of the differential operator $\rd_s$ 
and the shift operator $e^{-\rd_s}$.  As our derivation shows, 
its exponential $e^{\frakL}$ coincides with the first Lax operator $L$ 
of the 2D Toda hierarchy.  In other words, $\frakL$ is equal 
to $\log L$.  Emergence of the logarithm of a Lax operator 
is not a quite new circumstance.  Such operators are used 
for the construction of variants of the Toda lattice 
\cite{CDZ04,Getzler04}.  

The second half of this paper is devoted to a more systematic 
explanation of these observations in the language 
of {\it generalized string equations\/}.  
Generalized string equations are algebraic equations 
satisfied by the Lax and Orlov-Schulman operators 
$L,\bar{L},M,\bar{M}$ of the 2D Toda hierarchy \cite{TT95}. 
Generalized string equations for the tau function 
of the double Hurwitz numbers are presented 
in our previous work \cite{Takasaki12}.  
We show that those equations can be converted 
to a logarithmic form that contains $\log L$ and $\log\bar{L}$ 
rather than $L$ and $\bar{L}$.  The Lax operator $\frakL$ 
of the continuous Bogoyavlensky-Itoh hierarchy can be 
readily derived from these {\it logarithmic string equations\/}.

\section{Generating functions of Hurwitz numbers}

The (disconnected) Hurwitz numbers $H_d(\mu^{(1)},\ldots,\mu^{(r)})$ 
of $\CC\PP^1$ are defined by the sums 
\beqnn
  H_d(\mu^{(1)},\ldots,\mu^{(r)}) 
  = \sum_{[\pi]}\frac{1}{|\mathrm{Aut}(\pi)|}
\eeqnn
over all topological types $[\pi]$ of $d$-fold coverings 
$\pi: C \to \CC\PP^1$ of $\CC\PP^1$ 
by possibly disconnected compact Riemann surfaces $C$ 
with the ramification profile $(\mu^{(1)},\ldots,\mu^{(r)})$. 
The coverings are assumed to be ramified over $r$ points, 
say, $P_1,\ldots,P_r$, of $\CC\PP^1$. 
$\mu^{(j)}$'s are partitions of $d$, i.e., 
$\mu^{(j)} = (\mu^{(j)}_1,\mu^{(j)}_2,\ldots)$, 
$|\mu^{(j)}| = \sum_{i\geq 1}\mu^{(j)}_i = d$, 
and the $i$-th part $\mu^{(j)}_i$ of $\mu^{(j)}$ represent 
the order of cyclic ramification at the $i$-th point 
of $\pi^{-1}(P_j)$. $|\mathrm{Aut}(\pi)|$ denotes the number 
of covering automorphisms of $\pi$.  

Okounkov's ``double Hurwitz numbers'' are Hurwitz numbers 
of the form $H_d(\mu,\bar{\mu},1^{d-2}2,\ldots,1^{d-2}2)$, 
where $\mu$ and $\bar{\mu}$ are arbitrary partitions of $d$ 
and $1^{d-2}2$ denotes the partition $(2,1,\ldots,1)$. 
Let us use two sets of variables $\bsx = (x_1,x_2,\ldots)$, 
$\bar{\bsx} = (\bar{x}_1,\bar{x}_2,\ldots)$ 
and two parameters $\beta,Q$ to construct the following 
generating function of the double Hurwitz numbers: 
\beq
  z(\bsx,\bar{\bsx}) 
  = \sum_{r=0}^\infty \sum_{d=0}^\infty \sum_{|\mu|=|\bar{\mu}|=d}
    H_d(\mu,\bar{\mu},\underbrace{1^{d-2}2,\cdots,1^{d-2}2}_r) 
    \frac{\beta^r}{r!}Q^d p_\mu\bar{p}_\mu. 
  \label{z(x,xbar)}
\eeq
$p_\mu$, $\mu = (\mu_1,\mu_2,\ldots)$, and $\bar{p}_{\bar{\mu}}$, 
$\bar{\mu} = (\bar{\mu}_1,\bar{\mu}_2,\ldots)$,  
are the products $p_\mu = p_{\mu_1}p_{\mu_2}\cdots$, 
$\bar{p}_{\bar{\mu}} = \bar{p}_{\bar{\mu}_1}\bar{p}_{\bar{\mu}_2}\cdots$ 
of the power sums $p_k = \sum_{i\ge 1}x_i^k$, 
$\bar{p}_k = \sum_{i\ge 1}\bar{x}_i^k$. 
As pointed out by Okounkov \cite{Okounkov00}, 
one can rewrite this generating function as 
\beq
  z(\bsx,\bar{\bsx}) 
  = \sum_{\lambda\in\calP}e^{\beta\kappa(\lambda)/2}Q^{|\lambda|}
    s_\lambda(\bsx)s_\lambda(\bar{\bsx}), 
\eeq
where $\calP$ denotes the set of all partitions 
$\lambda = (\lambda_1,\lambda_2,\ldots)$ of arbitrary length, 
$\kappa(\lambda)$ is defined as 
\beqnn
  \kappa(\lambda) 
  = \sum_{i\geq 1}\lambda_i(\lambda_i-2i+1), 
\eeqnn
and $s_\lambda(\bsx)$ and $s_\lambda(\bar{\bsx})$ 
are the Schur functions in the sense of 
Macdonald's book \cite{Mac-book}.  

$z(\bsx,\bar{\bsx})$ corresponds to the tau function 
\beq
  Z(\bst,\bar{\bst}) 
    = \sum_{\lambda\in\calP}e^{\beta\kappa(\lambda)/2}Q^{|\lambda|}
    S_\lambda(\bst)S_\lambda(-\bar{\bst}) 
  \label{Z(t,tbar)}
\eeq
of the 2-component KP hierarchy with time variables 
$\bst = (t_1,t_2,\cdots)$ and 
$\bar{\bst} = (\bar{t}_1,\bar{t}_2,\cdots)$ 
by the transformations 
\beqnn
  t_k = p_k/k, \quad \bar{t}_k = - \bar{p}_k/k
\eeqnn
of the variables \cite{Okounkov00}. 
$S_\lambda(\bst)$'s are defined by the determinant formula 
\beqnn
  S_\lambda(\bst) = \det\left(S_{\lambda_i-i+j}(\bst)\right)_{i,j=1}^N,   
\eeqnn
where $N$ is chosen to be greater than or equal 
to the length of $\lambda$.  $S_n(\bst)$'s are defined 
by the generating function
\beqnn
  \sum_{n=0}^\infty S_n(\bst)z^n = \exp\left(\sum_{k=1}^\infty t_kz^k\right). 
\eeqnn

Specializing $Z(\bst,\bar{\bst})$ to $\bar{\bst} = (-c,0,0,\ldots)$ 
yields a generating function of the ``single Hurwitz numbers'' 
$H_d(\mu,1^{d-2}2,\ldots,1^{d-2}2)$ \cite{Okounkov00}. 
Note that $S_\lambda(-\bar{\bst})$ thereby turns into the special value 
\beq
  S_\lambda(c,0,0,\ldots) 
  = \frac{\dim\lambda}{|\lambda|!}c^{|\lambda|}, 
  \label{Schur(c)}
\eeq
where $\dim\lambda$ denotes the dimension 
of the irreducible representation of the symmetric group 
determined by $\lambda$.  In the following, 
the one-dimensional subspace $\bar{\bst} = (-c,0,0,\ldots)$ 
of the $\bar{\bst}$-flows is referred to as 
{\it the single Hurwitz sector\/}. 

$Z(\bst,\bar{\bst})$ can be extended to depend 
on a discrete variable $s \in \ZZ$ as 
\footnote{This generating function is slightly modified 
from our previous definition \cite{Takasaki12}.}
\beq
  Z(s,\bst,\bar{\bst}) 
  = \sum_{\lambda\in\calP}e^{\beta(\kappa(\lambda)+2s|\lambda|+(4s^3-s)/12)/2}
      Q^{|\lambda|+s(s+1)/2}S_\lambda(\bst)S_\lambda(-\bar{\bst}). 
\label{Z(s,t,tbar)}
\eeq
The deformations 
\beqnn
  \kappa(\lambda) \to \kappa(\lambda) + 2s|\lambda| + (4s^3-s)/12,\quad 
  |\lambda| \to |\lambda| + s(s+1)/2 
\eeqnn
of $\kappa(\lambda)$ and $|\lambda|$ stem from 
the matrix elements of operators $K$ and $L_0$ 
in a fermionic Fock space. 
$Z(s,\bst,\bar{\bst})$ thereby turns out to be a tau function 
of the 2D Toda hierarchy \cite{Okounkov00,Takasaki12}.  
Its specialization $Z(s,\bst,-c,0,0,\ldots)$ 
to the single Hurwitz sector is a tau function 
of the lattice KP (aka modified KP) hierarchy.

\section{Lax equations in single Hurwitz sector}

Our consideration is now focussed on the single Hurwitz sector 
$\bar{\bst} = (\bar{t}_1,0,0,\ldots)$.  
It is convenient to reorganize the expression (\ref{Z(s,t,tbar)}) 
of $Z(s,\bst,\bar{\bst})$ as 
\beq
  Z(s,\bst,\bar{\bst}) 
  = e^{\beta(4s^3-s)/24}Q^{s(s+1)/2}\tilde{Z}(s,\bst,\bar{\bst}), 
\eeq
where 
\beqnn
  \tilde{Z}(s,\bst,\bar{\bst}) 
  = \sum_{\lambda\in\calP}e^{\beta\kappa(\lambda)/2}
     (Qe^{\beta s})^{|\lambda|}S_\lambda(\bst)S_\lambda(-\bar{\bst}). 
\eeqnn
By the formula (\ref{Schur(c)}) of the special value 
of the Schur functions, $\tilde{Z}(s,\bst,\bar{\bst})$ 
in the single Hurwitz sector can be expressed as
\beq
  \tilde{Z}(s,\bst,\bar{t}_1,0,0,\ldots) 
  = \sum_{\lambda\in\calP}\frac{\dim\lambda}{|\lambda|!} 
     e^{\beta\kappa(\lambda)/2}(-Qe^{\beta s}\bar{t}_1)^{|\lambda|}S_\lambda(\bst). 
  \label{Ztd(s,t,tbar1)}
\eeq
Let $Z(s,\bst,\bar{t}_1)$ and $\tilde{Z}(s,\bst,\bar{t}_1)$ 
denote these specializations of $Z(s,\bst,\bar{\bst})$ 
and $\tilde{Z}(s,\bst,\bar{\bst})$.  

The foregoing expression of $Z(s,\bst,\bar{\bst})$ and 
its specialization $Z(s,\bst,\bar{t}_1)$ suggests 
to extend the range of $s$ from $\ZZ$ to $\RR$.  
In such an interpretation, $\tilde{Z}(s,\bst,\bar{t}_1)$ 
satisfies the differential equation 
\beq
  \frac{\rd\tilde{Z}(s,\bst,\bar{t}_1)}{\rd s} 
  = \beta\bar{t}_1\frac{\rd\tilde{Z}(s,\bst,\bar{t}_1)}{\rd\bar{t}_1} 
  \label{tbar1s-lineq}
\eeq
because this function depends on $s$ and $\bar{t}_1$ 
in such a form as $e^{\beta s}\bar{t}_1$. 
Let us consider implications of this fact. 

Let $\Psi(s,\bst,\bar{t}_1,z)$ denotes the Baker-Akhiezer function 
\beqnn
\begin{gathered}
  \Psi(s,\bst,\bar{t}_1,z) 
  = \frac{Z(s-1,\bst-[z^{-1}],\bar{t}_1)}
    {Z(s-1,\bst,\bar{t}_1)}z^se^{\xi(\bst,z)},\\
  [x] = (x,x^2/2,\ldots,x^k/k,\ldots),\quad 
  \xi(\bst,z) = \sum_{k=1}^\infty t_kz^k. 
\end{gathered}
\eeqnn
This function satisfies the auxiliary linear equations 
\beq
  \left(\rd_{t_k} - B_k\right)\Psi = 0,\quad k = 1,2,\ldots,\quad 
  \left(\rd_{\bar{t}_1} - \bar{u}_0e^{-\rd_s}\right)\Psi = 0 
  \label{single-lineq}
\eeq
of the positive and first negative flows in the 2D Toda hierarchy.  
$B_k$'s are difference operators of the form 
\beqnn
  B_k = (L^k)_{\ge 0} = e^{k\rd_s} + b_{k1}e^{(k-1)\rd_s} + \cdots + b_{kk}. 
\eeqnn
$L$ is the first Lax operator 
\beqnn
  L = e^{\rd_s} + u_1 + u_2e^{-\rd_s} + \cdots 
\eeqnn
of the 2D Toda hierarchy, and $(\quad)_{\ge 0}$ means 
extracting the non-negative powers of $e^{\rd_s}$.  
$\bar{u}_0$ is the function 
\beqnn
  \bar{u}_0 = \frac{Z(s,\bst,\bar{t}_1)Z(s-2,\bst,\bar{t}_1)}
              {Z(s-1,\bst,\bar{t}_1)^2}
\eeqnn
that arises in the leading part of the second Lax operator 
$\bar{L}$ as 
\beqnn
  \bar{L}^{-1} = \bar{u}_0e^{-\rd_s} + \bar{u}_1 
    + \bar{u}_2e^{\rd_s} + \cdots. 
\eeqnn

The equation with respect to $\bar{t}_1$ in (\ref{single-lineq}) 
can be converted to a bilinear differential equation 
for $Z(s,\bst,\bar{t}_1)$. One can rewrite 
$\bar{t}_1$-derivatives therein to $s$-derivatives 
with the aid of (\ref{tbar1s-lineq}). 
After some algebra, this bilinear differential equation 
turns into the linear equation 
\beq
  \left(\rd_s - \beta\bar{t}_1\bar{u}_0e^{-\rd_s}\right)\Psi 
  = (\log z)\Psi 
  \label{single-lineq2}
\eeq
for $\Psi(s,\bst,\bar{t}_1)$.  

We thus encounter a new Lax operator of the form 
\beq
  \frakL = \rd_s - ve^{-\rd_s}, \quad 
  v = \beta\bar{t}_1\bar{u}_0. 
  \label{fkL}
\eeq
As a consequence of (\ref{single-lineq2}) and the other equations 
of (\ref{single-lineq}), $\frakL$ satisfies the Lax equations 
\beq
  \frac{\rd\frakL}{\rd t_k} = [B_k,\frakL], \quad k = 1,2,\ldots. 
  \label{fkL-Laxeq}
\eeq
Moreover, since (\ref{single-lineq2}) implies 
the exponentiated equation $e^{\frakL}\Psi = z\Psi$ 
and $L$ satisfies the equation $L\Psi = z\Psi$, 
one can conclude that 
\beq
  e^{\frakL} = L. 
  \label{exp(fkL)=L}
\eeq
The reduced Lax operator $\frakL$ thus turns out to be 
the logarithm of $L$.  $L$, in turn, satisfies the Lax equations 
\beqnn
  \frac{\rd L}{\rd t_k} = [B_k, L]
\eeqnn
of the lattice KP hierarchy.  

Let us examine the lowest equation of (\ref{fkL-Laxeq}): 
\beq
  \left[\rd_{t_1} - e^{\rd_s}- u_1,\, \rd_s - ve^{-\rd_s}\right] = 0. 
\eeq
Upon substituting 
\beqnn
  u_1(s) = \frac{\rd\phi(s)}{\rd t_1},\quad 
  v(s) = e^{\phi(s)-\phi(s-1)},\quad 
  \phi(s) = \phi(s,\bst,\bar{t}_1), 
\eeqnn
this equation turns into the Toda-like field equation 
\beq
  \frac{\rd^2\phi(s)}{\rd t_1 \rd s}
  + e^{\phi(s+1)-\phi(s)} - e^{\phi(s)-\phi(s-1)} = 0. 
\eeq
This is exactly the continuum version 
of the Bogoyavlensky-Itoh equations \cite{Bog88,Itoh88}.  
We can thus reproduce the remark of 
Okounkov and Pandharipande \cite{OP01}.

\section{Logarithmic string equations}

Let us return to the double Hurwitz numbers, 
and consider the associated Lax operators $L,\bar{L}$ 
and the Orlov-Schulman operators 
\beqnn
\begin{gathered}
  M = \sum_{k=1}^\infty kt_kL^k + s + \sum_{n=1}^\infty v_nL^{-n},\\
  \bar{M} = - \sum_{k=1}^\infty k\bar{t}_k\bar{L}^{-k} 
    + s + \sum_{n=1}^\infty\bar{v}_n\bar{L}^n 
\end{gathered}
\eeqnn
of the full 2D Toda hierarchy \cite{TT95}. 
These operators are defined as 
\beqnn
\begin{gathered}
  L = We^{\rd_s}W^{-1},\quad 
  \bar{L} = \bar{W}e^{\rd_s}\bar{W}^{-1},\\
  M = W\left(s + \sum_{k=1}^\infty kt_ke^{k\rd_s}\right)W^{-1},\quad 
  \bar{M} 
  = \bar{W}\left(s - \sum_{k=1}^\infty k\bar{t}_ke^{-k\rd_s}\right)\bar{W}^{-1}
\end{gathered}
\eeqnn
by the dressing operators 
\beqnn
  W = 1 + \sum_{n=1}^\infty w_ne^{-n\rd_s},\quad 
  \bar{W} = \sum_{n=0}^\infty\bar{w}_ne^{n\rd_s}, 
\eeqnn
and satisfy the Lax equations 
\beq
\begin{gathered}
  \frac{\rd L}{\rd t_k} = [B_k,L],\quad 
  \frac{\rd L}{\rd\bar{t}_k} = [\bar{B}_k,L],\\
  \frac{\rd\bar{L}}{\rd t_k} = [B_k,\bar{L}],\quad 
  \frac{\rd\bar{L}}{\rd\bar{t}_k} = [\bar{B}_k,\bar{L}],
\end{gathered}
\label{Toda-Laxeq}
\eeq
where $B_k$'s are the same as those in (\ref{single-lineq}), 
and $\bar{B}_k$'s are difference operators of the form 
\beqnn
  \bar{B}_k = (\bar{L}^{-k})_{<0} 
  = \bar{b}_{k0}e^{-k\rd_s} + \cdots + \bar{b}_{k,k-1}e^{-\rd_s}, 
\eeqnn
namely, the operators obtained by extracting 
the negative powers of $e^{\rd_s}$ from $\bar{L}^{-k}$. 

We observed in our previous work \cite{Takasaki12} 
that these operators for the double Hurwitz numbers 
satisfy the generalized string equations 
\footnote{These equations are slightly different 
from those in the previous work due to modification 
of the tau function.}
\beq
  L = Qe^{\beta\bar{M}}\bar{L},\quad 
  \bar{L}^{-1} = QL^{-1}e^{\beta M}. 
  \label{gstreq}
\eeq
We here derive a logarithmic form of these equations, 
namely, equations for the logarithmic Lax operators. 
\beqnn
  \log L = W\rd_s W^{-1},\quad 
  \log\bar{L} = \bar{W}\rd_s\bar{W}^{-1} 
\eeqnn
and the Orlov-Schulman operators.  

A clue is the canonical commutation relations 
\beq
  [\log L, M] = 1,\quad [\log\bar{L}, \bar{M}] = 1. 
\eeq
One can use these relations and the Baker-Campbell-Hausdorff formula 
to rewrite the right sides of (\ref{gstreq}) as 
\beqnn
\begin{gathered}
  Qe^{\beta\bar{M}}\bar{L} = Qe^{\beta\bar{M}}e^{\log\bar{L}} 
   = \exp(\beta\bar{M} + \log\bar{L} - \beta/2 + \log Q), \\
  QL^{-1}e^{\beta M} = Qe^{-\log L}e^{\beta M} 
   = \exp(- \log L + \beta M - \beta/2 + \log Q). 
\end{gathered}
\eeqnn
Equating these results with the logarithm of the left sides 
of (\ref{gstreq}) yields {\it the logarithmic string equations\/}
\beq
\begin{gathered}
  \log L = \beta\bar{M} + \log\bar{L} - \beta/2 + \log Q,\\
  \log\bar{L} = \log L - \beta M - \beta/2 - \log Q 
\end{gathered}
\label{log-gstreq}
\eeq
for $\log L$, $\log\bar{L}$, $M$ and $\bar{M}$.  

This is not a perfect proof of these equations, 
because taking the logarithm of both sides of (\ref{gstreq}) 
can leave ambiguity of integral multiples of $2\pi\sqrt{-1}$.  
Actually, this ambiguity can be resolved by computations 
of the initial values of the Lax and Orlov-Schulman operators 
at a particular point of the $(\bst,\bar{\bst})$ space.  
We consider this issue in the next section.  

The reduced Lax operator (\ref{fkL}) in the single Hurwitz sector 
can be derived from (\ref{log-gstreq}) as well.  
Note that the first equation of (\ref{log-gstreq}) 
implies the relation 
\beqnn
  (\log L)_{<0} = (\beta\bar{M})_{<0} 
    = - \beta\sum_{k=1}^\infty k\bar{t}_k\left(\bar{L}^{-k}\right)_{<0}. 
\eeqnn
In the single Hurwitz sector $\bst = (\bar{t}_1,0,0,\ldots)$, 
this relation reduces to 
\beqnn
  (\log L)_{<0} = - \beta\bar{t}_1\bar{u}_0e^{-\rd_s}, 
\eeqnn
hence 
\beq
  \log L = \rd_s - \beta\bar{t}_1\bar{u}_0e^{-\rd_s}. 
\eeq
This is exactly the reduced Lax operator (\ref{fkL}).

\section{Perspective from factorization problem}

The generating function $Z(s,\bst,\bar{\bst})$ is derived 
from a fermionic formula \cite{Okounkov00,Takasaki12}. 
The fermionic construction of a tau function can be translated 
to the matrix factorization problem \cite{Takasaki84}
\beq
  \exp\left(\sum_{k=1}^\infty t_k\Lambda^k\right)U
  \exp\left(- \sum_{k=1}^\infty\bar{t}_k\Lambda^{-k}\right) 
  = W(\bst,\bar{\bst})^{-1}\bar{W}(\bst,\bar{\bst}), 
  \label{F-problem}
\eeq
where $U$ is the $\ZZ\times\ZZ$ matrix representing an element 
of the Clifford group $\widehat{\mathrm{GL}}(\infty)$ 
in the fermionic construction, 
$\Lambda^k$'s are the shift matrices $(\delta_{i+k,j})_{i,j\in\ZZ}$, 
and $W(\bst,\bar{\bst})$ and $\bar{W}(\bst,\bar{\bst})$ 
are lower and upper triangular matrices corresponding 
to the dressing operators.  
In the case of the double Hurwitz numbers, 
$U$ is a matrix of the form 
\beq
  U = e^{\beta(\Delta-1/2)^2/2}Q^{\Delta}, 
  \label{U-hurwitz}
\eeq
where $\Delta$ is the diagonal matrix $(i\delta_{ij})_{i,j\in\ZZ}$.
Note that $\Lambda$ and $\Delta$ are matrix representation 
of the difference operators $e^{\rd_s}$ and $s$ 
on the lattice $\ZZ$.  

We now extend this interpretation to the continuum $\RR$. 
Namely, $\Lambda$ and $s$ are understood to be 
the difference operators 
\beq
  \Lambda = e^{\rd_s}, \quad \Delta = s 
\eeq
defined on $\RR$.  (\ref{F-problem}) thereby becomes 
a factorization problem for difference operators. 

Since (\ref{U-hurwitz}) is a diagonal matrix, 
the factorization problem can be solved explicitly 
at the particular point 
\beqnn
  \bst = \bszero,\quad 
  \bar{\bst} = - \bsc = (-c_1, -c_2, \ldots),
\eeqnn
where $c_k$'s are arbitrary constants. The solutions, 
which can be identified with the initial values 
$W_0 = W(\bszero,-\bsc)$, $\bar{W}_0 = \bar{W}(\bszero,-\bsc)$ 
of the dressing operators, read 
\beq
\begin{gathered}
  W_0 = e^{\beta(s-1/2)^2/2}Q^s
        \exp\left(- \sum_{k=1}^\infty c_ke^{k\rd_s}\right)
        Q^{-s}e^{-\beta(s-1/2)^2/2},\\
  \bar{W}_0 = e^{\beta(s-1/2)^2/2}Q^s. 
\end{gathered}
\eeq

These expressions of $W_0$ and $\bar{W}_0$ enable us 
to compute the associated initial values 
\beqnn
\begin{gathered}
  \log L_0 = L(\bszero,-\bsc) = W_0\rd_sW^{-1},\quad
  \log\bar{L}_0 = \bar{L}(\bszero,-\bsc) = \bar{W}_0\rd_s\bar{W}_0^{-1},\\
  M_0 = M(\bszero,-\bsc) = W_0 sW_0^{-1},\quad 
  \bar{M}_0 = \bar{M}(\bszero,-\bsc) 
    = \bar{W}_0\left(s + \sum_{k=1}^\infty kc_ke^{-k\rd_s}\right)\bar{W}_0^{-1}
\end{gathered}
\eeqnn
of the logarithmic Lax operators and the Orlov-Schulman operators. 
The outcome takes the following form: 
\beq
\begin{gathered}
  \log L_0 = \rd_s 
    + \beta\sum_{k=1}^\infty kc_kQ^ke^{-\beta k(k+1)/2}e^{k\beta s}e^{-k\rd_s},\\
  \log\bar{L}_0 = \rd_s - \beta(s- 1/2) - \log Q,\\
  M_0 = \bar{M}_0 
  = s + \sum_{k=1}^\infty kc_kQ^ke^{-\beta k(k+1)/2}e^{k\beta s}e^{-k\rd_s}. 
\end{gathered}
\eeq
This implies the algebraic relations 
\beq
\begin{gathered}
  \log L_0 = \beta\bar{M}_0 + \log\bar{L}_0 - \beta/2 + \log Q,\\
  \log\bar{L}_0 = \log L_0 - \beta M_0 - \beta/2 - \log Q, 
\end{gathered}
\eeq
namely, the logarithmic string equations (\ref{log-gstreq}) 
are satisfied at the initial time $\bst = \bszero$, $\bar{\bst} = - \bsc$.  
This is enough to conclude that (\ref{log-gstreq}) themselves 
are satisfied, because both sides of these equations 
solve the same Lax equations as (\ref{Toda-Laxeq}), 
and one can resort to the uniqueness of the initial value problem.

\section{Conclusion}

The Bogoyavlensky-Itoh hierarchies \cite{Bog88,Itoh88,Bog87,Itoh87} 
are variants of the well known Volterra lattice.  
We have shown that the continuum version \cite{Bog88,Itoh88} 
of these integrable hierarchies underlies the single Hurwitz numbers.  
In this respect, recent work of Dubrovin et al. 
on cubic Hodge integrals \cite{DLYZ1612,LZZ17} is very interesting.  
They proved that the Volterra lattice is an integrable structure 
of cubic Hodge integrals in a special case \cite{DLYZ1612}, 
and conjectured a similar link with variants 
of the Volterra lattice in more general cases \cite{LZZ17}.  
We can show, by the factorization technique of Section 5, 
that the finite-step version \cite{Bog87,Itoh87} 
of the Bogoyavlensky-Itoh hierarchies 
is indeed hidden in those cubic Hodge integrals. 
This issue will be reported elsewhere.

\subsection*{Acknowledgements}

This work is partly supported by the JSPS Kakenhi Grant 
JP25400111 and JP18K03350.


\begin{thebibliography}{99}


\bibitem{Okounkov00}
A.~Okounkov, 
Toda equations for Hurwitz numbers, 
Math. Res. Lett. {\bf 7} (2000), 447--453. 

\bibitem{GJ08}
I.~P.~Goulden and D.~M.~Jackson, 
The KP hierarchy, branched covers, and triangulations, 
Adv. Math. {\bf 219} (2008), 932--951. 

\bibitem{MM08}
A.~Mironov and A.~Morozov, 
Virasoro constraints for Kontsevich-Hurwitz partition function, 
JHEP {\bf 0902} (2009), 024. 

\bibitem{Kazarian08}
M.~Kazarian, 
KP hierarchy for Hodge integrals, 
Adv. Math. {\bf 221} (2009), 1--21. 

\bibitem{Alexandrov11}
A.~Alexandrov, 
Matrix models for random partitions, 
Nucl. Phys. {\bf B851} (2011) 620--650. 

\bibitem{Takasaki12}
K.~Takasaki,
Generalized string equations for double Hurwitz numbers, 
J. Geom. Phys. {\bf 6}2 (2012), 1135--1156. 

\bibitem{AMMN12}
A.~Alexandrov, A.~Mironov, A.~Morozov and S.~Natanzon, 
Integrability of Hurwitz partition functions. I. Summary, 
J. Phys. A: Math. Theor. {\bf 45} (2012), 045209.

\bibitem{AMMN14}
A.~Alexandrov, A.~Mironov, A.~Morozov and S.~Natanzon, 
On KP-integrable Hurwitz functions, 
JHEP {\bf 11} (2014), 080. 

\bibitem{HO15}
J.~Harnad and A.~Yu.~Orlov, 
Hypergeometric $\tau$-functions, Hurwitz numbers 
and enumeration of paths, 
Comm. Math. Phys. {\bf 338} (2015), 267--284. 

\bibitem{Bog88}
O.~I.~Bogoyavlensky, 
The Lax representation with a spectral parameter 
for certain dynamical systems, 
Izv. Akad. Nauk SSSR Ser. Mat. {\bf 52} (1988), Issue 2, 243–-266. 

\bibitem{Itoh88}
Y.~Itoh, 
Integrals of a Lotka-Volterra system of infinite species, 
Progr. Theoret. Phys. {\bf 80} (1988) 749--751. 

\bibitem{Bog87}
O.~I.~Bogoyavlensky, 
Some constructions of integrable dynamical systems, 
Izv. Akad. Nauk SSSR Ser. Mat. {\bf 51} (1987), Issue 4, 737--766. 

\bibitem{Itoh87}
Y.~Itoh, 
Integrals of a Lotka-Volterra system 
of odd number of variables, 
Progr. Theoret. Phys. {\bf 78} (1987), 507--510. 

\bibitem{OP01}
A.~Okounkov and R.~Pandharipande, 
Gromov-Witten theory, Hurwitz numbers, and matrix models, I, 
arXiv:math/0101147. 

\bibitem{CDZ04}
G.~Carlet, B.~Dubrovin and Y.~Zhang, 
The extended Toda hierarchy,
Moscow Math. J. {\bf 4} (2004), 313--332. 

\bibitem{Getzler04}
E.~Getzler, 
The equivariant Toda lattice, 
Publ. RIMS, Kyoto Univ., {\bf 40} (2004), 507--536. 

\bibitem{TT95}
K.~Takasaki and T.~Takebe, 
Integrable hierarchies and dispersionless limit, 
Rev. Math. Phys. {\bf 7} (1995), 743--808.  

\bibitem{Mac-book}
I.~G.~Macdonald, 
{\it Symmetric functions and Hall polynomials\/}, 
Oxford University Press, 1995.

\bibitem{Takasaki84}
K.~Takasaki,
Initial value problem for the Toda lattice hierarchy, 
Adv. Stud. Pure Math. vol. 4, Mathematical Society of Japan, Tokyo, 1984, 
pp. 139--163. 

\bibitem{DLYZ1612}
B.~Dubrovin, S.-Q.~Liu, D.~Yang and Y.~Zhang, 
Hodge-GUE correspondence and the discrete KdV equation, 
arXiv:1612.02333. 

\bibitem{LZZ17}
S.-Q.~Liu, Y.~Zhang and C.~Zhou, 
Fractional Volterra hierarchy, 
Lett. Math. Phys. {\bf 108} (2018), 261--283. 

\end{thebibliography}
\end{document}